**Title:**

# Using a complex system approach to address world challenges in Food and Agriculture.

**Author names and affiliations:**
H.G.J. van Mil[a], E.A. Foegeding[b], E.J. Windhab[c], N. Perrot[d], E. van der Linden[a,e]

[a] TI Food and Nutrition, Nieuwe Kanaal 9A, 6709 PA Wageningen, The Netherlands, har.science@gmail.com, vanderlinden@tifn.nl

[b] Department of Food, Bioprocessing, and Nutrition Sciences, North Carolina State University, Box 7624, Raleigh, NC 27695-7624, United States.
e-mail: eaf@unity.ncsu.edu

[c] Laboratory of Food Process Engineering, Institute of Food, Nutrition and Health, Department of Health Science and Technology, ETH Zurich, Schmelzbergstrasse 9, 8092 Zurich, Switzerland.
e-mail: erich.windhab@hest.ethz.ch

[d] UMR782 Génie et Microbiologie des Procédés Alimentaires, AgroParisTech, INRA, 78850 Thiverval-Grignon, France.
e-mail: nathalie.perrot@grignon.inra.fr

[e] Laboratory of Physics and Physical Chemistry of Foods, Wageningen University and Research Center, Bornse Weilanden 9 (building 118) 6708 WG Wageningen, The Netherlands.
e-mail: erik.vanderlinden@wur.nl

**Corresponding author.**
E. van der Linden. e-mail: erik.vanderlinden@wur.nl Phone: +31 317 485515

**Abstract**
World food supply is crucial to the well-being of every human on the planet in the basic sense that we need food to live. It also has a profound impact on the world economy, international trade and global political stability. Furthermore, consumption of certain types and amounts foods can affect health, and the choice of livestock and plants for food production can impact sustainable use of global resources. There are communities where insufficient food causes nutritional deficiencies, and at the same time other communities eating too much food leading to obesity and accompanying diseases. These aspects reflect the utmost importance of agricultural production and conversion of commodities to food products. Moreover, all factors contributing to the food supply are interdependent, and they are an integrative part of the continuously changing, adaptive and interdependent systems in the world around us. The properties of such interdependent systems usually cannot be inferred from the properties of its parts. In addressing current challenges, like the apparent incongruences of obesity and hunger, we have to account for the complex interdependencies among areas such as physics and sociology. This is possible using the complex system approach. It encompasses an integrative multi-scale and inter-disciplinary approach.
Using a complex system approach that accounts for the needs of stakeholders in the agriculture and food domain, and determines which research programs will enable these stakeholders to better anticipate emerging developments in the world around them, will enable them to determine effective intervention strategies to simultaneously optimise and safeguard their interests and the interests of the environment.

# 1. Introduction
The current and future challenges for the agriculture and food (Agri&Food) area reside in continuously providing safe, tasty, healthy, affordable, and sustainably produced food in sufficient amounts. Food itself, its harvesting, separation, production, marketing, innovations, and other aspects, can be considered as parts of a continuously changing, self-organising, interactive and adaptive complex system. Understanding such complex systems, and their interdependencies, is needed in order to meet current and future Agri&Food challenges. As professor Allen has put it (Allen, 2012): "*Systems evolve qualitatively over time and interact with each other. For example, the human body has adapted to food and in turn the food has been adapted to it. The food we consume is governed by culture and lifestyle, but also by climate, soil and energy issues. All these aspects are in a continuous state of flux*". The systems we refer to are also known as complex systems. The science that aims to understand such systems is known as the Science of Complexity. This science acknowledges the necessity of an integrative approach. It also acknowledges the fact that (Allen, 2012) "*one needs to be aware that our predictions do not necessarily come true*", implying that the assumption of a reductionist approach needs to be reconsidered. An important practical challenge is more accurate predictions on the basis of limited and widely dispersed information(Jaynes, 2003). Taking into account the inherent uncertainties in complex systems while considering problems to be solved, an achievable goal is to articulate strategies towards solutions, instead of formulating the solution.

The Science of Complexity has vast applications beyond Agri&Food due to its inherent design in addressing superficially non-connected phenomena. During the last century, the public became better educated, informed and aware about its own place within the food network and the challenges it raises for the sustainability of human life on earth. The general perception that our current way of life is not sustainable motivates actions on both local and global scales in different societal, political and ecological arenas. In the resulting societal and political discourse, the breadth of the spatial and temporal scales frustrates decision-making and action on global and long timescales due to the sheer complexity of the problem, which puts strain on our cognitive skills and our psychological inability to deal with long-term solutions and rewards. This adds to the importance and relevance of the Science of Complexity.

The current paper originates from the outcome of a symposium held in Oosterbeek, The Netherlands, and on subsequent work. The symposium entitled "Agri-Food and Science of Complexity" was held in June 2012, under sponsorship of the Top Institute Food and Nutrition in The Netherlands. In section 2 we describe some characteristics of complex systems. In section 3 we discuss important spatial scales in more detail. In section 4 we list methodologies that are applied for analysis of complex systems. In section 5 we give some Agri-Food topics amenable to a science of complexity approach. We end with conclusions and perspectives on several worldwide challenges in Agri-Food.

## 2. Complex systems and some important aspects of our approach

A complex system can be defined as a system that consists of parts that are interrelated and from which one cannot infer the behaviour of that system. A measure of complexity can be defined as "the amount of information necessary to describe the system" (Bar-Yam, 1997).

Many typical characteristics of a complex system can be mentioned (Bar-Yam 1997). One is the existence of different spatial scales where the overall system behaviour needs a multi-scale description. For a human being for example, the smallest relevant scale may be the molecular scale, at which molecules assemble into structures that together form the larger scale, i.e. that of the cell. The cells in turn are assembled into organs, muscles and bones, which together are assembled into parts of the human body. The behaviour at the scale of a cell influences behaviour on the larger scale of an organ, and reversely, stresses at the scale of the animal affect processes on the smaller scale of the cell. The properties of the cell require a different terminology than the properties of the entire human. Many different (meta-) stable states are possible. Interdependencies within one system occur because of interactions among the parts that exist at the same or different spatial scales within the system.

There are three main aspects that are important to the complexity approach.
The first is how one describes the specific problem or topic in terms of its conceptual level and scale of observation (spatial and temporal).
The second is to identify the amount of information necessary to describe the problem or topic. This amount of information is dependent on the scale (detail) of observation of the system for which the problem is described. For example, describing the random motion of gas molecules requires a lot of information. This information can be greatly reduced when describing phenomena at a macroscopic scale, say pressure, by means of the use of only a few thermodynamic parameters. Increasing the scale of the observer simplifies the description, i.e. it decreases the amount of information required to describe the system at that scale. This is referred to as the so-called *complexity profile* (Bar-Yam 1997). This profile helps to simplify the description of a complex system. Refinements to this approach have been put forward recently(Harmon, & Bar-yam, 2012).
The third important aspect is to estimate the effectiveness of interventions. Effective intervention might call for orchestrated actions between different spatial scales to obtain a desired and robust result. In order to define a robust strategy for action, one needs to identify those spatial levels that effectively (as in the sustainability issue) affect the desired outcome. To be effective, those levels need to allow for some level of controlled intervention. To allow for such complex orchestrated interventions, a variety of disciplines that are based on fundamentally different conceptual frameworks need to work together; this calls for an interdisciplinary input into the complex problem. The effectiveness of an intervention is thus not only related to the level of control, but also to the conceptual strength of the relevant academic discipline and how it can incorporate results from other, sometimes conceptually disjoint, disciplines of the other spatial scales. A good measure for conceptual strength can be defined in terms of information theory; the least complex and most relative informative model is to be preferred (Bar-Yam, 2005). Developing models at the different spatial scales and combining them into one larger model to allow for the orchestrated action is the best strategy. The control at different relevant spatial scales and integrated interpretations of the relevant conceptual levels are key to an effective approach. The impact of the possible interventions/policies should be studied, or simulated, and according refinements regarding the relevant spatial scale and complexity should be made in subsequent cycles.
Regarding the effectiveness of interventions in complex systems, a "one solution fits all" interventions/policies will not be effective, but, instead, multi-level interventions that are congruent with the complexity of the problem are required (Bar-Yam 2004). If however the problem is not complex, a "simple solution fits all" may well apply. For example, the problem of high number of people having goiter could be solved by adding iodine to table salt (practiced e.g. in the US from 1924 onwards). The problem for anybody who had goiter was deficiency of iodine, and the simple solution, for everybody, was adding iodine, by means of adding it to a food ingredient that everybody consumes (table salt). Similarly, the lack of vitamin C causes scurvy, which can be easily cured by administering vitamin C. In contrast, a phenomenon like obesity is caused by many different factors at the same time, which are not always the same for everybody (for an overview of factors see (MURPHY, 1960), for ingredient effects (Meydani & Hasan, 2010), and for oral processing effects (Rolls, 2011)). A solution to problems like obesity requires multi-level interventions and policies.

## 3. Scientific disciplines

The following sub-sections provide general information on the scientific disciplines in relation to our approach. It should be noted that these are broad classifications used to illustrate the concept of a complex system approach and has the potential for further refinement.

**3.1 Material sciences**
Starting at the material science level, in the lower left part of figure 1, food can be seen as the result of a process where biological molecules assemble into food structures over a period of time, t, as a function of molecular composition and concentration, c, externally applied stresses E, and internal stresses/molecular mobility. Stress in this case can be of different natures, e.g. mechanical or electromagnetic.
This 4 dimensional representation is somewhat similar to the idea of principal axes in a jamming state diagram as introduced by Liu and Nagel (Liu & Nagel, 1998). The two axes, on external stresses and internal stresses, reflect a distinction between the food system and its environment. We note that from an engineering point of view, it would be more natural to consider applied power per volume instead of applied energy per volume (i.e. stress) but in view of symmetry with respect to the axis for internal stress/mobility we have chosen for stress. The resulting structure is composed of different levels. Starting with individual molecules at the nano-scale, they are assembled into a range of structures that are described at the meso-scale. The meso-scale structures similarly contribute to the macrostructure. All or only specific structural levels are important for attributes describing food products. For example, aroma intensity profiles are molecule based, while graininess (a textural attribute) is based on the mesostructure and macrostructure level (one only perceives grains larger than around 20-50 micrometer). Following Crutchfield (Crutchfield & Machta, 2011; Crutchfield, 2011), this meso-structure embodies the historic information that has been stored within the specific food as it has co-evolved over its specific trajectory throughout the diagram until time t. As such, food structure is a reflection of the historic information acquired, i.e. encompassing the specific route that has led to the food. Recall that in this context, *historical* refers to the time course of growing or directly assembling molecules into the food (the latter also often referred to as "processing" the food).

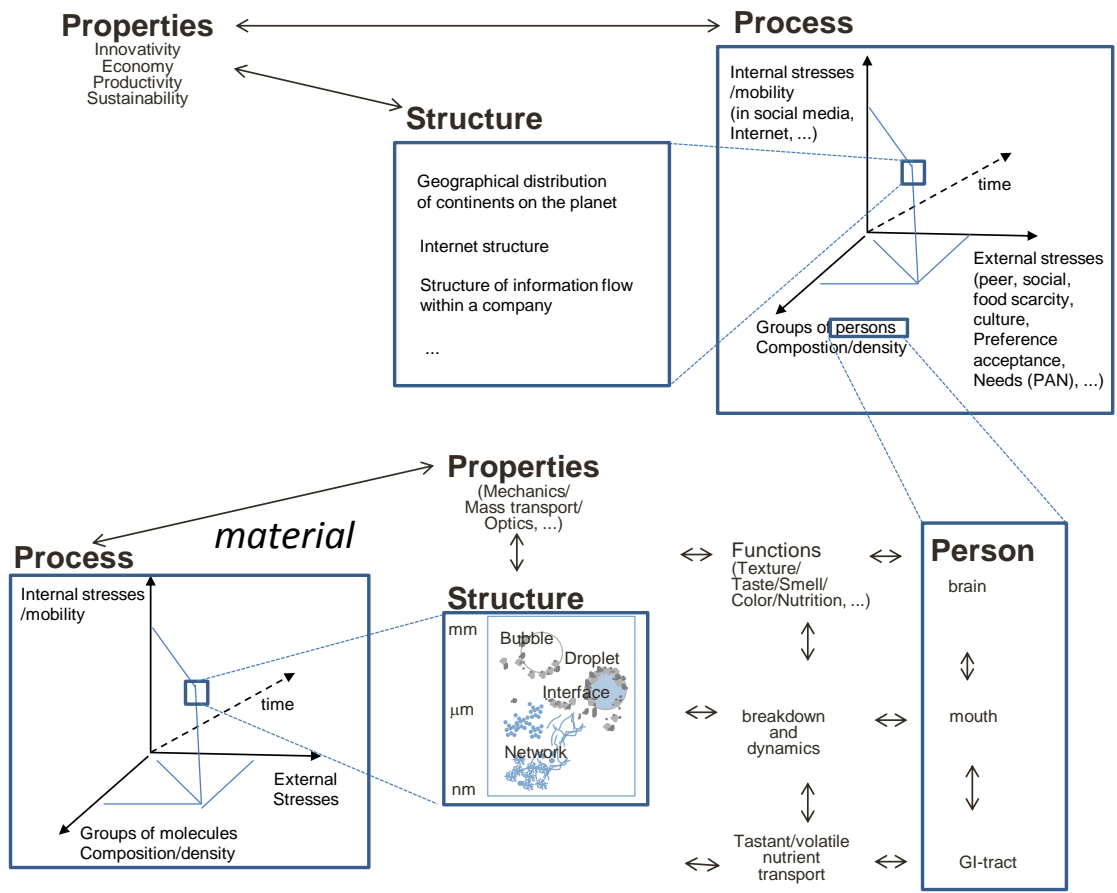

*Figure 1. Processes leading to structures and properties at various scales within an Agri-Food context.*

The specific route is defined by the type of ingredient(s), their concentration c, time t, externally applied stresses, E, and internal stress/mobility. This is the case for processed food, such as bread, as well as for foods such as fruits and vegetables where the structures are formed by biological processes during growth and ripening. Thus, we arrive at a triangle of interrelationships between structure, properties and processing. The importance of such triangular interrelationships has been put forward and extensively addressed by one of us (Windhab 2008). The effect of ingredient type, and the (in-)ability to have them interchanged without affecting food properties has also been subject of many papers (see e.g. a paper on the replacement of egg white protein with milk proteins in making angel food cake (Berry, Yang, & Foegeding, 2009; Pernell, Luck, Foegeding, & Daubert, 2002; Yang & Foegeding, 2011). Structure is important for many different fields, such as in designing complex chemical systems using nature inspired approaches (Coppens, 2012). For foods, structural changes are also important while the food it is being stored, transported, or displayed in shops which can effects their organoleptic properties.

### 3.2 Life Sciences
The life sciences come into play in the selection, evolution and annual growth of plants and animals determining the composition and distribution of food; the formation of plant and animal materials into food structures; and the breakdown and assimilation of food.

The breakdown and assimilation involve the mixture of desirability-related traits (appearance, flavour, taste and texture) that initiate consumption; and the processes occurring in the gastrointestinal track that determine delivery of nutrients and bioactive compounds. These are related to the macro- and meso-structure dynamics (during mastication, swallowing, and digestion). As such, material properties are related to sensory attributes, and nutrient delivery.

For sensory attributes, we distinguish texture (mechanics), fracture and breakdown during mastication (mechanics and sound), colour (optics), taste (tastant release and transport to sensors) and smell (volatile release and transport to sensors). As a complicating factor in sensory perception, the senses are known to mutually influence one another, due to the rules that the brain uses to integrate and interpret information. For instance, certain sensory attributes of a product perceived via one or more modalities (such as vision and touch) can bias a consumer's perception of other attributes of that product derived from other sensory modalities into alignment, and consequently, modulate a person's overall (multisensory) consumption experience. (Spence, 2012; Piqueras-Fiszman & Spence, 2012).

Once the food enters the gastrointestinal tract, the food starts to co-evolve with the bio-system present there. The dynamics, adaptability and self-organisation occurring in this environment is staggering. It is remarkable that one still manages to deduce guiding principles that describe the essentials of the system behaviour (Barbara M. Bakker, Karen van Eunen, Jeroen A.L. Jeneson, Natal A.W. van Riel, Frank J. Bruggeman, 2010; Teusink, Westerhoff, & Bruggeman, 2010).

### 3.3 Behavioural Sciences
The physical and chemical properties of food determine what is sensed by physiological processes but the sensing will often be modified by previously stored information. An interesting coupling exists between structure and behaviour during mastication. The structure may be broken down during mastication in a specific way, leading to a specific perception of texture. Reversely, this perception determines how we masticate (behaviour). Molecular absorption during consumption and digestion can also influence eating behaviour, both on a short and longer-term basis. Food liking and resulting buying behaviour are normative reflections based on a complex mix of personal preference within a particular demographic and cultural background; these are partially genetically determined (Bartoshuk, 2000) and by the social interactions within a group and between groups of people. This gets us to the society level.

### 3.4 Social Sciences
At a social level, the preference and liking of groups of persons will depend on their personal preferences, their health, as well as their history and interactions with their peers in that group (culture). Groups (aging population, patients, children…), in turn, can show interactions with each other. All these interactions have social dimensions, conditions alluding to food security, economic and sustainability aspects, to name a few.

For the purpose of modelling, human beings are often considered as agents, analogous to molecules in a material. This leads to an interesting analogy between the social science level and the material science level. For the social level one dimension is formed by the typical properties of each individual agent, as well as the composition and density within that group of agents (analogous to type and concentration of molecules). Another dimension at the social level is formed by culture, food scarcity, peer pressure, social pressure, and so-called PAN variables (Preference, Acceptance, Needs (Windhab 2008)). These factors are the analogue of externally applied stresses at the material level. Yet another dimension on the social level is formed by internal relations and personal mobility within the group (the analogue of internal stresses/mobility at the material level). The analogy extends itself to an emerging social structure that embodies the historic information that is stored within the social system, as it has co-evolved along the trajectory in the diagram. Also on this level the amount of historic information that a system stores may be considered a measure of the (structural) complexity of that system (Crutchfield, 2012). Furthermore, analogous to the material point of view, the emerging social structures on this level are determining how the social systems process and store information, i.e. how they are determining the system properties. One of the interesting and complex features at this level is that the actions of the agent at this higher level (i.e. of the human being) have cognitive and psychophysical determinants that are different in risky and non-risky contexts (Tversky & Kahneman, 1983), and the agents have judgements that are subject to heuristics and biases (Tvenky & Kabneman D., 1974).

Other aspects at the social level are e.g. of an economic nature, including product development processes and innovation processes (Bonabeau & Meyer, 2001; Frenken, 2006). Some recently developed tools to address and picture emerging structures at that level are described in the literature (Chavalarias & Cointet, 2008)An interesting example on optimisation of cheese ripening has been described, which includes aspects on quantitative material science and qualitative knowledge of process operators (Barrière et al., 2013; Baudrit, Sicard, Wuillemin, & Perrot, 2010; Perrot, Trelea, Baudrit, Trystram, & Bourgine, 2011)The role of computer science and the description of digital systems for business conduction is nicely addressed (Razavi, Moschoyiannis, & Krause, 2009). In regards to issues of economy and food scarcity, some important aspects are addressed recently (Lagi, Bertrand, & Bar-yam, 2011). In regards to a living system example, we like to mention ecology of fish populations (Allen and McGlade, 1987).

## 4. Methodologies

Machine learning methodologies (Chavalarias & Cointet, 2008; Lutton et al., 2011) could be an integral part of a complex systems approach to analysing problems in the Agric Food area. Models for communications and information distribution can follow existing lines (Allen & McGlade, 1987; Bonabeau & Meyer, 2001; Frenken, 2006). One may mention:

1. Methods developed in systems biology (Teusink et al., 2010)and behavioural research (Piqueras-Fiszman, Velasco, & Spence, 2012; Spence, Harrar, & Piqueras-Fiszman, 2012) are key due to their complex and integrated designs.
2. Computational methods for analyzing the resulting complex datasets are akin to general algorithms from machine learning and Graphical models (Chavalarias & Cointet, 2008). New developments in artificial intelligence and evolutionary algorithm and interactive learning are also interesting approaches that allow capturing implicit expert knowledge (Lutton et al., 2011). Some applications of such methodologies have already been applied on food science (Perrot 2011).
3. Scaling relations need to be given for an reinterpretation at the different levels. Here input is needed from the relevant disciplines. An example is the Structure, Process and Properties (S-PRO$^2$) method (Windhab 2008) constructing a modular model for direct implementation.
4. A theoretical approach can be used in deriving relations from different information sources. This method can help with the interpretation within the different levels.
5. The formal methods with logical or mathematical structure can be extended by directed data-searches. The actually mappings are the domain of mathematics.

The key to success in inter-disciplinary projects is the completeness of the research network in terms of knowledge and the efficiency of the flow of information through the network (Allen & McGlade, 1987; Frenken, 2006). The problem of completeness can be solved by powerful methods based on semantic analysis (Chavalarias & Cointet, 2008). The resulting research network, as defined by these algorithms, can be mapped into the procedure described above.

Once the network is realized, a model is developed that is based on multidisciplinary cooperation. The multi-level nature of complex problems, involving everything from molecular to social interactions, inherently requires interdisciplinarity. Researchers must extent their working knowledge beyond their own discipline to achieve the required knowledge and semantic overlap among the disciplines; in other words, the network needs to learn. Some models and conditions for innovation have been discussed in the context of Swarm intelligence (Bonabeau & Meyer, 2001)and in terms of agent communication languages (Razavi et al., 2009)

In addition to these methodologies, *gamification* might also be of interest. In gamification, methods developed in the area of recreational games are implemented in non-game contexts. These methods can be implemented in a cooperation research model where different actors bring their own expertise into the projects akin to the so called Foldit game (Hescher, 2012). The inferential power of this gamification method of humans is attenuated by inferential power of computers by giving more attention to the human - computer cooperation (Sankar, 2012). Data visualization, akin to the methods developed in bio-informatics and economics (H Rosling, 2011), come naturally with the machine learning approaches and will be important within the game context.

## 5. Topics in Agri&Food amenable to science of complexity

Academia and industry together have identified various project topics on their various conceptual levels in a Food Innovation Summit organized by the Top institute Food and Nutrition (TIFN) in Oosterbeek, the Netherlands in 2012, see figure 2. We note that the topic of sustainability has been the topic of the Food Innovation summit 2013, also organized by TIFN in Oosterbeek, The Netherlands.

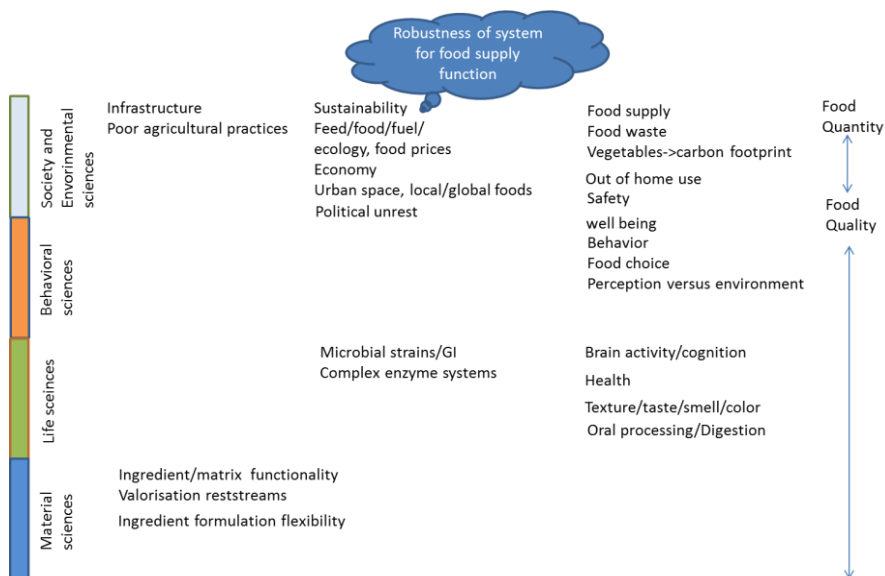

*Figure 2. Project topics and their conceptual levels as jointly identified by academia and industry in the TIFN Food innovation Summit 2012.*

## 6. Systematic approach to complex problems.

Having identified the topics as in figure 2, the first step for action is to identify a problem/question within that topic. For example: "The level of obesity increases in country x and what intervention strategies can one employ to decrease or reverse this decrease?" Then we are confronted with two tasks.

One is the harvesting of the information within each level. For this there are many different methods available and used (MacKay, 2003). The other deals with the fact that there is not one discipline that is able to research the whole range of levels mentioned above. However we can circumvent these barriers by carefully organizing the information flow by explicitly mapping relevant bits of information to disciplines higher in the organizational level. Although abstract methods from the theoretical sciences like mathematics and logic are used at all hierarchical levels, they lack any interpretation of their own on the subject matter. Mathematics and logic are mere instruments in this context at the level of syntax, the important semantics is left to the respective disciplines. Considering this second task, there is an example developed in physics of how to connect different time and or length scales by using a scaling approach (Barenblatt, 2003). There are also examples using scaling approaches in the other levels such as life and behavioural sciences (Bettencourt, Lobo, Strumsky, & West, 2010; Kello et al., 2010)

We subsequently have to systematically integrate the two tasks into one methodology. The methodology is based on the following roadmap of procedures (see figure 3):

1. Formulating a well-posed research question, at a particular level, sets the context which defines the conditional structure; its operational form as a study design where lower hierarchal levels are nested in the higher levels. In figure 3 we give an example for four different levels (represented by different sciences).
2. Modern methods in statistics (MacKay, 2003) allow for the analysis of these complex and compound nested data-set. Here the field of machine learning and Bayesian algorithms provide key methods. These methods can do the "bookkeeping" of the project as they particularly focus on the conditional structure. Power laws and dimensions of the leading variables, will serve as output for these algorithms. These power laws are derived inductively.
3. The interpretation of the power laws at the relevant aggregation levels is expressed in scaling laws. These laws identify relations that can span a number of levels and are applied in almost every relevant discipline; they are expressed in the form of powers of dimensions relevant to that aggregation level. Scaling laws however, are not always easy to find or interpretable in the different levels. These scaling laws are derived via deductively.
4. Scaling laws are indications for special types of underlying structures. They are well known in physics (Barenblatt, 2003) and engineering. They can be linked to underlying principles of control, robustness and self-organisation (Csete & Doyle, 2009) and conservation laws.
5. When these underlying principles have been identified, the resulting structures can be investigated in relation to each other; this can be done conceptually or by using mathematics at a more abstract level. These relations can be seen as mapping of one structure into the other. It is the inverse problem of the nested design in procedural step 1 above. This step leads to a generic model linking all hierarchal levels to each other.
6. This last result in 5 can be used in formulating a new research question, which brings you again to procedural step 1.

For application and product development steps 1 to 3 are sufficient. In order to obtain generalizations, steps 4 and 5 are necessary.

To summarize our approach, an inductive empirical description sets the conditional and correlation structure of the problem. This is followed by a deductive theoretical description of the underlying structure.

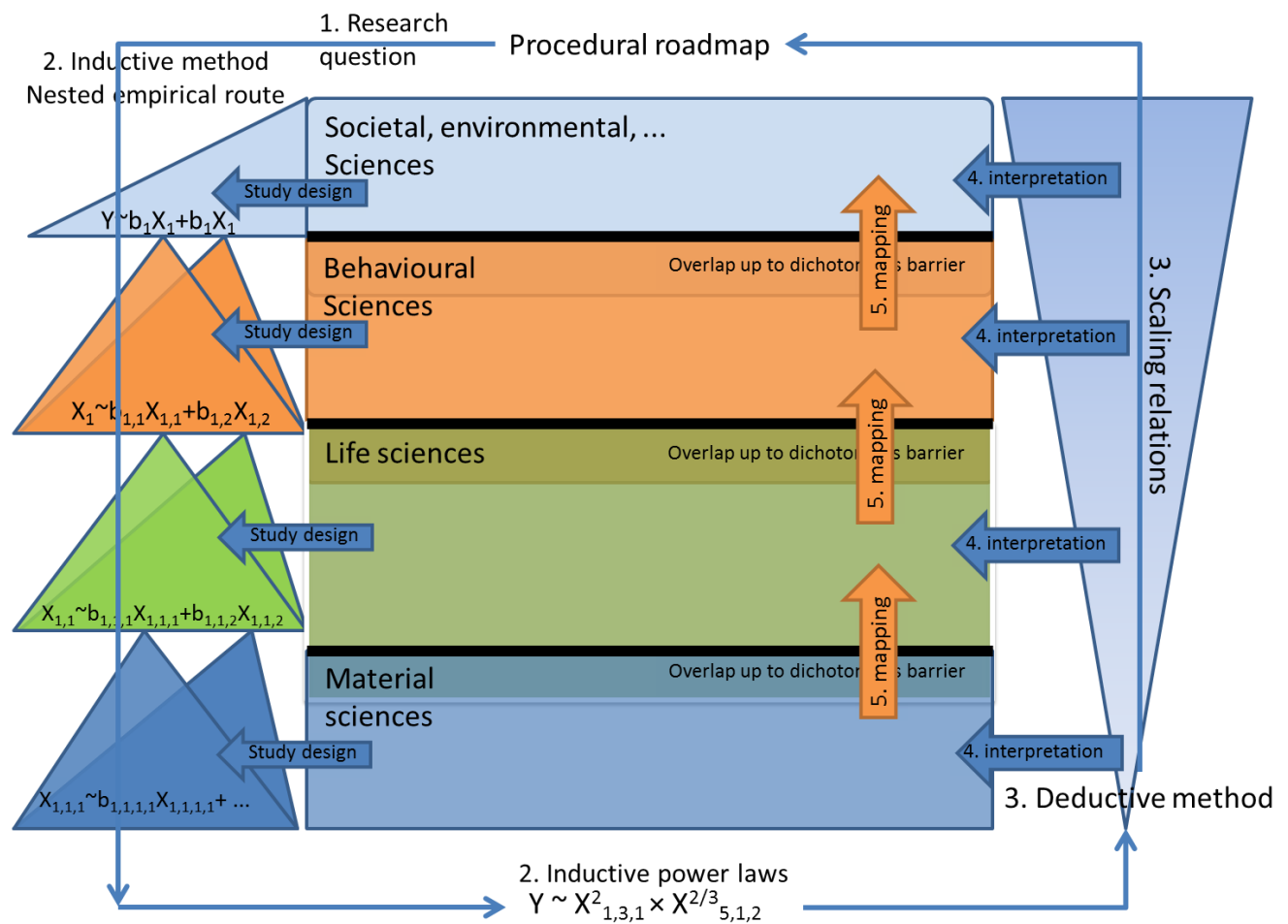

Figure 3. Inductive and deductive methods to address specific research questions requiring a *complex system approach*

## 7. Perspectives

Usually, in linking different hierarchical levels and their respective disciplines, one is hampered by the fact that these levels are conceptually disjoint, leading to dichotomies. The combined effort of the proposed approaches tries to rectify this problem.

Usually, at the highest hierarchical level, we expect more than one class of correct answers. For instance, individual persons make their decisions based on a weighing of complex factors. This implies that one should not aim for identifying "the most probable state" but, instead, accept the existence of different distributions in different populations, and the existence of individuals with different "life-histories".

To educate the next generation of scientist, academia should introduce a more integrated perspective on education. The recognition of students that research problems may be multifaceted and can be investigated using different perspectives and disciplines, that need to be integrated, is an important step. A new set of skills needs to be added to the student curriculum to effectively deal with the Argi&Food challenges of today and the future.

The combination of science of complexity and needs in the Agri&Food area will enable its stakeholders to effectively address future challenges while maintaining the capacity to endure.

## Acknowledgements.

We acknowledge the Top Institute of Food and Nutrition for financial support and organizing the Food Innovation Summit 2012 and for financial support later on (Harald van Mil, Erik van der Linden). We acknowledge discussions with all contributors during the summit and/or comments on the manuscript by Peter Allen, Yaneer Bar-Yam, Eric Bonabeau, David Chavalarias, Brian Guthrie, Marc-Olivier Coppens, Koen Frenken, Bruce German, Peter Krause, Theo Odijk, Marcel Paques, Betina Piqueras-Fiszman, Peter Schall, Charles Spence and Bas Teusink. EvdL gratefully acknowledges the inspiring discussions on the topic of complexity during the last 5 years with Theo Odijk.

## References.

Erich J. Windhab, Sep 30, 2008 – A Process Engineering Approach. Dialogue on. Food, Health and Society. Sept. 29 - 30, 2008. Swiss Re Centre
*http://www.zhaw.ch/fileadmin/user_upload/life_sciences/Dateien/News_Veranstaltungen/Tagungen/Dialogue_Food_Health_Society/Presentation_Erich_Windhab.pdf* . (date: 31-07-2013)